# Spectral Response Function Guided Deep Optimization-driven Network for Spectral Super-resolution

Jiang He, Jie Li*, *Member, IEEE,* Qiangqiang Yuan*, *Member, IEEE,* Huanfeng Shen, *Senior Member, IEEE,* and Liangpei Zhang, *Fellow, IEEE*

*Abstract*—**Hyperspectral images are crucial for many research works. Spectral super-resolution (SSR) is a method used to obtain high spatial resolution (HR) hyperspectral images from HR multispectral images. Traditional SSR methods include model-driven algorithms and deep learning. By unfolding a variational method, this paper proposes an optimization-driven convolutional neural network (CNN) with a deep spatial-spectral prior, resulting in physically interpretable networks. Unlike the fully data-driven CNN, auxiliary spectral response function (SRF) is utilized to guide CNNs to group the bands with spectral relevance. In addition, the channel attention module (CAM) and reformulated spectral angle mapper loss function are applied to achieve an effective reconstruction model. Finally, experiments on two types of datasets, including natural and remote sensing images, demonstrate the spectral enhancement effect of the proposed method. And the classification results on the remote sensing dataset also verified the validity of the information enhanced by the proposed method.**

## I. INTRODUCTION

**H**YPERSPECTRAL (HS) imaging is a technique used to explore the spectral characteristics of objects completely via the fine resolution of scene radiance. Hyperspectral images (HSIs) processing, such as segmentation [1], classification [2], detection [3], [4], and tracking [5], have gained increasing attention due to the rich spectral information. HS imaging has also been developed for numerous applications ranging from remote sensing [6]-[8] to medical imaging [9].

This work was supported by the National Natural Science Foundation of China, grant number 41701400.

J. He, J. Li, and Q. Yuan are with the School of Geodesy and Geomatics, Wuhan University, Hubei, 430079, China (e-mail: jiang_he@whu.edu.cn; jli89@sgg.whu.edu.cn; yqiang86@gmail.com).

H. Shen is with the School of Resource and Environmental Sciences, Wuhan University, Wuhan, Hubei, 430079, China (e-mail: shenhf@whu.edu.cn)

L. Zhang is with the State Key Laboratory of Information Engineering in Surveying, Mapping, and Remote Sensing, Wuhan University, Wuhan, Hubei, 430079, China (e-mail: zlp62@whu.edu.cn)

*Corresponding: jli89@sgg.whu.edu.cn; yqiang86@gmail.com.

Hyperspectral sensors acquire scene radiance with numerous spectral bands in a fine wavelength range. However, less energy radiance is sensed by each detector element when the spectral resolution is high. The sensors require long exposure time to obtain an acceptable signal-to-noise-ratio of each band. Compared with Red-Green-Blue (RGB) and multispectral images (MSIs), HSIs always lack fine spatial resolution. This limitation affects the availability of HSIs for applications that require high spatial resolution. Many researchers have proposed the direct reconstruction of HR HSIs by image super-resolution (SR) of low spatial-resolution (LR) HSIs to enhance the spatial details of HSIs. Akgun et al. [10] proposed a model that can represent the hyperspectral observations as weighted linear combinations and used a set-theoretic method as a solution. Gu et al. [11] proposed an SR algorithm that uses an indirect approach based on spectral unmixing and designed learning-based SR mapping as the backpropagation neural network. The aforementioned methods only utilize LR HSIs to reconstruct HR HSIs. However, poor spatial enhancement is observed when the ratio between LR and HR is large.

With the development of detector elements, abundant sensors are currently designed to achieve a good representation of spatial details and temporal variations. However, these sensors capture only three or four spectral bands for a very high spatial resolution ($\leq$10 m), especially for remote sensing satellites, such as Sentinel-2, GaoFen-2, QuickBird, and WorldView. Although MSIs generally have a high spatial resolution, they cannot completely represent the spectral characteristics of the object by using only a few spectral channels.

Combining the respective advantages of HSIs and MSIs, some researchers use HR MSIs as auxiliary data to improve the spatial resolution of HSIs. Hardie et al. [12] presented a novel maximum a posteriori (MAP) estimator for enhancing the spatial resolution. The MAP estimator used a spatially varying statistical model based on vector quantization to exploit localized correlations. Kawakami et al. [13] fused HSIs with images from RGB cameras by initially applying an unmixing algorithm to the hyperspectral input and then regarding the unmixing problem as the search for input factorization. In [14], Akhtar et al. proposed a fusion algorithm of MSIs and HSIs using non-parametric Bayesian sparse representation. Meng and Zhang et al. [15] proposed an integrated relationship model that relates to the HSIs and multi-source HR observations based on the MAP framework. Palsson et al. [16] proposed a novel method for the fusion of MSIs and HSIs, which is performed in the low-dimensional PC subspace; thus, only the first few PCs must be estimated instead of all spectral bands. The fusion-based method can substantially improve the spatial resolution of the image through the HR spatial detail injection. However, the HR MSIs corresponding to the LR HSIs covering the same area and acquired at a similar time are not always easily accessible in many cases. Although HR MSI data were available, the registration and preprocessing of multi-sensor data are difficult. Besides, this difficulty affects the accuracy and performance of algorithms.

The SSR methods are proposed to overcome the unavailability of HRHS images by increasing the spectral resolution of MS images without auxiliary HS images, which focuses on the spectral transformation rather than the spatial resolution enhancement. In 2008, Parmar et al. [17] first reconstructed HS image from RGB image by sparse recovery. Inspired by this research, Arad, and Ben-Shahar [18] proposed the computation of the dictionary representation of each RGB pixel by using the orthogonal match pursuit algorithm. Wu et al. [19] substantially improved Arad's method by pretraining an overcomplete dictionary as anchor points to perform a nearest neighbor search based on the A+ algorithm proposed by Timofte et al. from spatial SR [20]. In 2018, Akhtar et al. [21] modeled natural spectra under Gaussian processes and combined them with RGB images to recover HS images. Without dictionary learning, Nguyen et al. [22] explored a strategy to train a radial basis function network that presents the spectral transformation to recover the scene reflectance using training images. Deep learning, especially CNN, has recently attracted increasing attention and been demonstrated to outperform most traditional approaches in areas, such as segmentation [23], classification [24], denoising [25], and spatial SR [26]. Inspired by the semantic segmentation architecture *Tiramisu* [27], Galliani et al. [28] proposed *DenseUnet* with 56 convolutional layers to show good performance. To prove that comparable performance can be achieved by shallow learning, Can et al. [29] proposed a moderately deep residual CNN to recover spectral information of RGB images. Shi et al. [30] designed a deep CNN with dense blocks and a novel fusion scheme to deal with the situation when the spectral response function is unknown. Optimizing bands pixel by pixel, Gewali et al. [31] proposed a deep residual CNN to learn both the optimized MS bands and the transformation to reconstruct HS spectra from MS signals. Arun et al. [32] explored a CNN based encoding-decoding architecture to model the spatial-spectral prior to improve recovery. However, the deep learning-based model is similar to a data-driven black box with the ideal capability of feature learning and nonlinear mapping. Recently, interpretability specific to the problem has been identified as an important part of CNN development. Some research works have attempted to achieve this purpose. Most of them are trying to combine deep learning with physical model-driven methods. By learning a regularization term for the variational model or MAP framework, CNNs are utilized to achieve some physical mappings as approximate operator and denoiser in many image processing tasks, such as denoising [33], [34], compressive sensing [35], data fusion [36], and deblurring [37]. However, these methods just utilized the pre-trained CNN prior but did not update it in model-driven optimization. And the training of those algorithms is broken into two stages: learning optimization and variational optimization, which is difficult to inherit the data-driven advantages of deep learning.

In this paper, an end-to-end optimization-driven CNN with the spectral degradation model is built and different spectral ranges are grouped to be reconstructed based on spectral response functions. The spectral response function is utilized to guide the CNN group in the spectral similar bands to further enhance spectral information. Rather than alternately running a variational model and CNN, an optimization-driven CNN with deep spatial-spectral prior and parametric self-learning is proposed. The proposed CNN repeatedly updates the intermediate HS image in an end-to-end manner. The contributions are as follows.

1) An end-to-end optimization-driven CNN is proposed by combining the data-driven method with the optimization algorithm to improve the model interpretability. The channel attention module is introduced in the proposed model to embed the parameter self-learning considering spectral differences of bands into CNN.
2) The SRF is employed as a guide to aid CNN in grouping suitable spectral bands to reconstruct hyperspectral information and learn good spectral details from the true spectral channel ranges in the proposed CNN.
3) The spatial-spectral convolutional layers are used to model deep spatial-spectral prior. And the proposed network also employed a fast spatial-spectral loss function reformulated from L1 and spectral angle mapper losses to reach quick convergence and good spatial-spectral constraints.

The remaining part of the paper is organized as follows. Section II describes the degradation model and derives the spectral super-resolution algorithm based on the variational model to proposed optimization-driven CNN. Section III presents the experiments on two types of datasets, including five datasets from natural to remote sensing images, and some discussions of deep learning-based methods are also made. Finally, we draw some conclusions in section IV.

## II. PROPOSED METHOD

Firstly, the spectral degradation between MS and HS imaging is modeled in this section. Based on this model, the SSR problem is formulated and split into two subproblems. Finally, by learning physical mappings using CNNs, the proposed spectral SR network with a joint spatial-spectral HSI prior (HSRnet) is comprehensively demonstrated. The framework of the proposed method is illustrated in Fig. 1. The proposed framework can be divided into two parts, including an initial restoration network and optimization stages with attention-based parametric self-learning and spatial-spectral networks, which followed the data flow in model-based methods.

### A. Model Formulation

Let $X \in R^{W \times H \times C}$ represent the observed HSI, where $C$ is the number of the spectral channels, and $W$ and $H$ are the width and height, respectively. $Y \in R^{W \times H \times c}$ represents the observed multispectral image, where $c < C$ is the number of multispectral bands, specifically for RGB image, with $c = 3$. Varying in SRF, the sensors obtain different MS or HS data with different bands. A transformation matrix $\Phi \in R^{c \times C}$ can be used to describe the spectral degradation between MS and HS imaging as follows.

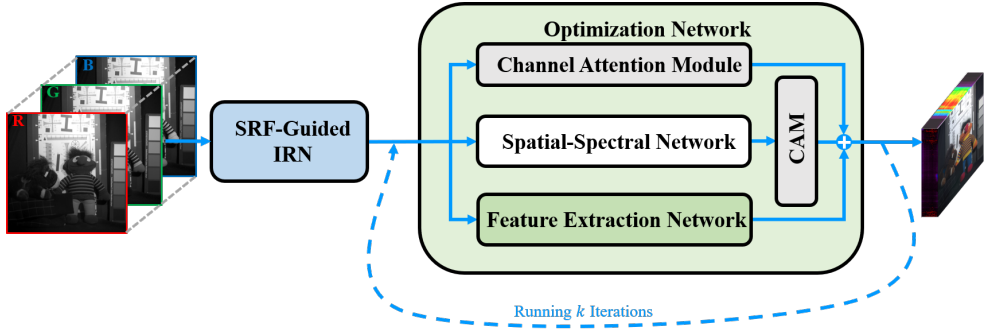

Fig. 1  Framework of the proposed HSRnet

$$Y = \Phi X \qquad (1)$$

The spectral transform matrix is closely related to SRF, which can be approximately estimated by some methods, such as Hysure [38] and RWL1-SF [39]. According to (1), the relationship between MSIs and HSIs is illuminated. However, in SSR, obtaining a high-dimension cube from low-dimension data is an under-determined problem. The high-dimension HSIs can be approximately predicted by adopting some priors to a minimization problem to constrain the solution space as follows:

$$\widehat{X} = arg\ min_{X} \|Y - \Phi X\|^2 + \gamma \mathcal{R}(X) \qquad (2)$$

where $\gamma$ is a trade-off parameter, and $\mathcal{R}(\cdot)$ is a regularization function. As in (2), the minimization problem is constrained by two parts. The first term is the data fidelity term that limits the solution according to the degradation model, and the second regularization term constrains the predicted $\widehat{X}$ with an HSI prior.

The variable splitting technique can be employed to further solve this minimization problem and separate the two terms in (2). An auxiliary variable $H$ is introduced to reformulate (2) to obtain a constrained optimization problem, which is shown as follows:

$$\widehat{X} = arg\ min_{X} \|Y - \Phi X\|^2 + \gamma \mathcal{R}(H),\ s.t.\ H = X \qquad (3)$$

According to the half-quadratic splitting method, the cost function is then transformed into

$$L(\widehat{X}, \widehat{H}) = \|Y - \Phi X\|^2 + \mu \|H - X\|^2 + \gamma \mathcal{R}(H) \qquad (4)$$

where $\mu$ is a penalty parameter with various values in different iterations. Using the variable splitting technique, Equation (4) can be resolved by solving two subproblems iteratively as

$$\widehat{X}^{k+1} = arg\ min_{X} \|Y - \Phi X\|^2 + \mu \|H^k - X\|^2 \qquad (5)$$

$$\widehat{H}^{k+1} = arg\ min_{H} \|H - X^{k+1}\|^2 + \lambda \mathcal{R}(H) \qquad (6)$$

where $\lambda = \gamma/\mu$ is another penalty parameter related to $\mu$ and $\gamma$. The degradation model $\Phi$ and HSI prior $\mathcal{R}(H)$ can be considered individually due to the variable splitting technique.

Considering the $X$-subproblem, instead of directly solving the $X$-subproblem as a least-squares problem, an approximate solution updated by the gradient descent algorithm is employed in this paper as follows:

$$\widehat{X}^{k+1} = X^k - \varepsilon[\Phi^T(\Phi X^k - Y) + \mu(\ X^k - H^k)]$$
$$= [(1 - \varepsilon\mu)I - \varepsilon\Phi^T\Phi]X^k + \varepsilon\Phi^T Y + \varepsilon\mu H^k \qquad (7)$$

As mentioned in [33], the $H$-subproblem in (6) can be rewritten as

$$\widehat{H}^{k+1} = arg\ min_{X} \frac{1}{2(\sqrt{\lambda/2})^2} \|H - X^{k+1}\|^2 + \mathcal{R}(H) \qquad (8)$$

Equation (8) can be regarded as denoising (both in spatial and spectral domain) images with the noise level of $\sqrt{\lambda/2}$ with the constraint of HSI priors. And the prior includes two meanings: one is the restraint on spatial information, for example, clearer edges, texture features, local smoothness, non-local self-similarity, and non-Gaussianity; the other is the restraint on spectral information, such as sparsity and high correlations between spectra. Unlike the total variation or sparsity prior, the HSI prior contains more than one property which should be modeled with nonlinearity to increases the accuracy [40].

With good nonlinear learning ability, deep learning-based methods are proved to be capable of many image restoration tasks. In this paper, a spatial-spectral network (SSN) is proposed to achieve the optimization as (8) describes because of the nonlinearity of HSI prior. By extracting spatial and spectral information, the intermediate results are updated following the constraint of (6). Thus, the optimization of $H$ is rewritten as

$$\widehat{H}^{k+1} = Spa\_Spec(X^k) \qquad (9)$$

where $Spa\_Spec(\cdot)$ presents the SSN. The details will be described in the later subsection. With a new way of updating $H$, the original optimization method, which alternatively updates $H$ and $X$ until convergence, can be rewritten to a unified updating of $X$. Considering (7) and (9), reformulated optimization is as follows:

$$\widehat{X}^{k+1} = \widetilde{\Phi}X^k + \varepsilon\Phi^T Y + \varepsilon\mu \cdot Spa\_Spec(X^k) \qquad (10)$$

where $\widetilde{\Phi} = (1 - \varepsilon\mu)I - \varepsilon\Phi^T\Phi$ indicates a new transformation matrix to the intermediately reconstructed image $X^k$.

With the help of the gradient descent algorithm and the HSI prior, the proposed method is to update the intermediate $X^k$ with a linear combination of three parts, including the initial restoration $\Phi^T Y$, the transformed $X^k$, and the spatial-spectral prior to $X^k$. The initial restoration $\Phi^T Y$, $\widetilde{\Phi}$, and parameters $\varepsilon$ and $\mu$ are also replaced with convolutional layers because the CNN has been employed to model the HSI prior, which is as follows:

$$\widehat{X}^{k+1} = T(X^k) + \varepsilon \cdot IRN(Y) + \varepsilon\mu \cdot Spa\_Spec(X^k) \qquad (11)$$

where $T(\cdot)$ presents the transformation layer of $X^k$. One convolutional layer is utilized in this paper. $IRN(\cdot)$ indicates the initial restoration network block. All parameters, namely, $\varepsilon$ and $\mu$, are learned by channel attention module. Details are presented later.

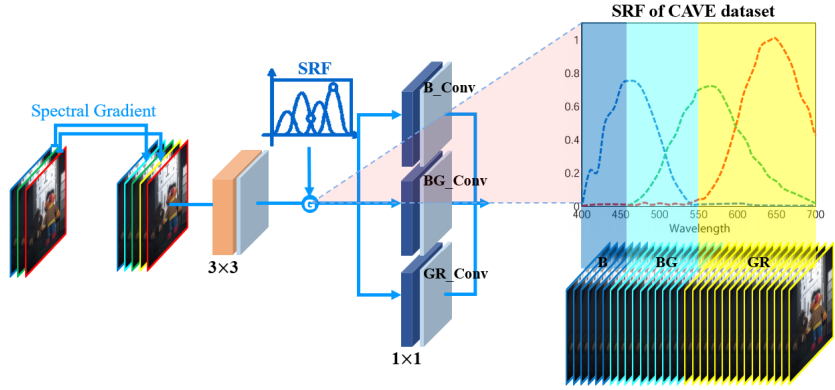

Fig. 2  IRN block

### B. SRF-Guided Initial Restoration

As described in Section I, the SRFs can provide spectral relevance between MS and HS bands from an imaging point of view. Therefore, unlike the traditional deep learning-based methods, SRF guiding is introduced as an auxiliary operation, which can realize effective SSR performance. Auxiliary physical operations give a great deal of assistance to deal with image restoration in many types of research [41]-[44]. In the proposed CNN, a new SRF-guided IRN block is proposed to group bands by spectral radiation characteristics and reconstruct the initial SSR result $X^0$ with different operators. The SRF-guided initial restoration network is shown in Fig. 2.

The whole block is a two-layer CNN. And the reconstruction convolutional layers for different spectral ranges are identified separately using SRF as a guide. Details are as follows. First, the spectral gradients of RGB/MS image are computed to construct a data cube with a dimension of $W \times H \times (2c-1)$ as shown in Fig. 3.

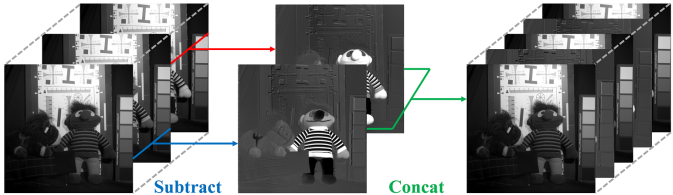

Fig. 3  Spectral gradient computation

After that, the data cube is fed into a $3 \times 3$ convolutional layer to extract spectral features. These features are then fed into SRF-guided convolutional layers by grouping with spectral relevance according to SRFs. The spectral grouping is used to avoid reconstruction distortion caused by the excessive spectral difference between different channels. By roughly representing spectral relevance from the similarity of imaging according to spectral response functions, SRF-guided convolutional layers don't have to be adjusted for the same sensor, which improves the generalization of this module.

For example, in CAVE dataset, which consists of RGB images and HSIs with 31 bands, spectral ranges can be divided into three classes, including only covered by the blue band, covered by blue and green bands, and covered by green and red bands, according to the spectral response function. Then the grouped spectral features are respectively fed into convolutional layers. So, SRF-guided convolutional layers play a role as spectral grouping restoration. In other words, HS channels with high spectral relevance will be constructed by the same convolution operator group.

With SRF as a guide, the IRN block can group the spectral bands with a high spectral correlation. This grouping avoids the introduction of irrelevant spectral information that disrupts spectral restoration.

### C. Deep Spatial-Spectral Prior

As discussed in Section IIA, the HSI prior can be modeled by a spatial-spectral network, which is shown in Fig. 4. The SSN comprises two subnetworks in series: one for spatial information extraction and the other for spectral feature extraction.

The intermediate reconstructed HSI is fed into the first $3 \times 3$ convolutional layer to compute for additional feature maps considering the influence of spatial neighborhood and transform the HSI data into a high-dimensional space. This transformation provides additional extracted features to the subsequent learning of spectral information. The second $3 \times 3$ convolutional layer is used as a selection for the next spectral optimization from the redundant features; besides, reducing the number of feature maps can accelerate the network calculation [45]. The last $1 \times 1$ convolutional layer achieves the fine-tuning of each spectral vector pixel by pixel. With the data-driven training, fine-tuning can be learned as spectral optimization processing. Furthermore, the $1 \times 1$ convolutional layer can significantly improve the effect of low-level image processing, which can further facilitate SSN learning of the HSI prior [46]. A skip connection adding the input to the output of the spatial network is also applied. This connection can accelerate network computation and simultaneously force the network to provide further attention to the changing details.

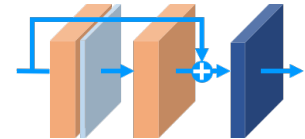

Fig. 4  Spatial-spectral network

Equipped with spatial-spectral networks, the proposed method can implicitly introduce the HSI prior to further constrain the solution space and achieve improved SSR results.

### D. Optimization Stages in HSRnet

With the application of the gradient descent algorithm and deep spatial-spectral prior, the SSR problem can be solved by updating $\boldsymbol{X}$ as (11), which is regarded as an optimization process. When the optimization is unfolded, a network comprising multiple stages can serve as an alternative to achieve optimization update in a deep-learning manner, as shown in the optimization stages in Fig. 5.

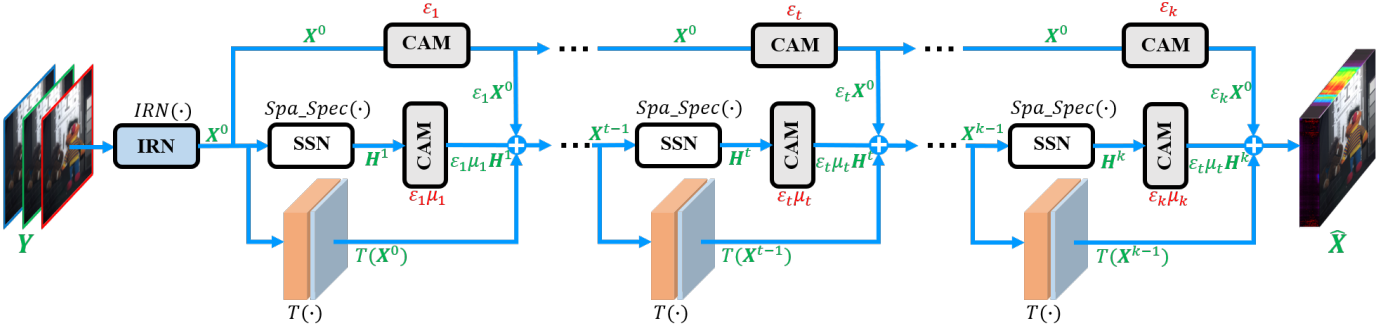

Fig. 5 The optimization stages of HSRnet

The original RGB/MS image $\boldsymbol{Y}$ is first fed into the IRN block for an initial estimation $\boldsymbol{X}^0 = IRN(\boldsymbol{Y})$. Given the initial HSI restoration $\boldsymbol{X}^0$, the iterative optimization, which can be trained to learn the HSI prior and match the spectral degradation model simultaneously, can be modeled in a feed-forward manner. Three parts are needed for the $k$th updating as shown in (11). The first term is $T(\boldsymbol{X}^{k-1})$, a spectral transformation preceding $\boldsymbol{X}^{k-1}$, which is computed by a convolutional layer with a size of $C \times 3 \times 3 \times C$. The second term is $\varepsilon \cdot IRN(\boldsymbol{Y})$, which is the weighted initial estimation $\boldsymbol{X}^0$ by $\varepsilon$. The last is $\varepsilon\mu \cdot Spa\_Spec(\boldsymbol{X}^{k-1})$, the $\varepsilon\mu$-weighted result of $\boldsymbol{H}^k$, which is the result from $\boldsymbol{X}^{k-1}$ fed into the SSN for the HSI prior. The parameters $\varepsilon$ and $\mu$ are learned by a block with attention mechanism. Details are provided later.

### E. Attention-Based Parametric Self-learning

The step size $\varepsilon$ and the balance parameter $\mu$ change accordingly in each iteration to optimize the intermediate variable $\boldsymbol{X}^k$ iteratively. All the parameters in this paper can be learned due to the backpropagation in training, which is a data-driven manner without manual intervention. However, parameters in traditional methods are all similar for different spectral channels. This similarity may be an inappropriate way for spectral bands with different radiance characteristics because of different optimal signal-to-noise ratios and different spectral information introduced in the input data. Considering the radiance differences in different bands and the good performance in the channel weighting of CAM, the CAM blocks are applied to the proposed HSRnet as shown in Fig. 6. CAM can help HSRnet focus on bands that need urgent optimization with high weights by exploiting the inter-channel relationship of features.

The CAM block comprises two pooling layers with max- and mean-pooling, two $3 \times 3$ convolutional layers, and a sigmoid function. First, the reconstructed HSI is fed into the pooling layer to extract global weights. After pooling layers, the global weights are forwarded into two convolutional layers and summed. Finally, the channel weights are activated by a sigmoid function before element-wise multiplication.

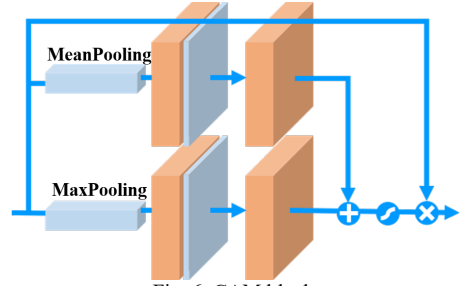

Fig. 6 CAM block

Introducing channel attention, HSRnet can easily learn different parameters as a vector of each iteration rather than a fixed value. This condition can ensure the adaptive weight adjustment of the network in spectral optimization and the realization of an improved reconstruction effect.

### F. Fast Joint Spatial-Spectral Loss

The L1 loss and spectral angle mapper (SAM) loss functions are applied in this paper as shown below to enhance spectral resolution and preserve the spatial detail simultaneously.

$$L = \left| \widehat{\boldsymbol{X}} - \boldsymbol{X} \right| + \alpha \sum_{j=1}^{WH} \cos^{-1}\left( \frac{\hat{X}^j X^j}{\sqrt{\hat{X}^{j^T} \hat{X}^j} \sqrt{X^{j^T} X^j}} \right) \quad (12)$$

where $\widehat{\boldsymbol{X}}$ is the reconstructed HSI, $\boldsymbol{X}$ is the ground truth, $\hat{X}^j$ presents the recovered spectral vector in $j$th pixel, $X^j$ is the ground truth, and $\alpha$ is a balance parameter. However, the application of SAM loss is difficult in practice due to computational complexity and the inability of GPU-accelerated computation as a vector form. Inspired by [47], a transformed RMSE loss is utilized as a substitute for SAM loss, which is shown as

$$L = \left| \widehat{\boldsymbol{X}} - \boldsymbol{X} \right| + \alpha \cos^{-1}\left( 1 - \tfrac{1}{2} \| \widehat{\boldsymbol{X}}' - \boldsymbol{X}' \|^2 \right) \quad (13)$$

where $\widehat{\boldsymbol{X}}'$ is the reconstructed HSI unitized pixel by pixel, and $\boldsymbol{X}'$ is the unitized ground truth.

TABLE I
RUNNING TIME OF DIFFERENT LOSSES

|  | With CPU | With GPU |
|---|---|---|
| SAM Loss | 2.6642 s | - |
| Proposed Loss | 1.4611s | 0.03748 s |

Thus, SAM loss can be calculated as a tensor form. This calculation allows parallel computation with GPU, which will be swift in learning as shown in Table I.

## III. EXPERIMENTAL RESULTS

### A. Experimental Setting

#### 1) Comparison Methods

The proposed method is compared with the related algorithms of SSR without HSI required as input, including Arad [18], A+ [19], DenseUnet [28], CanNet [29], HSCNN+ [30], and sRCNN [31]. The compared methods involve the dictionary and deep learning-based methods, which are currently state-of-the-art in SSR. The models of A+ and Arad are reproduced through a program [19] coded by Wu et al.

#### 2) Quantitative Metrics

Four quantitative image quality metrics, including correlation coefficient (CC), peak signal-to-noise ratio (PSNR), structural similarity (SSIM) [49], and spectral angle mapper (SAM) [50], are utilized to evaluate the performance of all comparison methods quantitatively. CC, PSNR, and SSIM are indexes that show the spatial fidelity of the reconstructed HSIs, which are computed on each channel and averaged over all spectral bands. Results with their large values indicate that the method is effective for maintaining spatial detail. Meanwhile, SAM evaluates the spectral preservation of the algorithms, showing improved spectral fidelity when the SAM is small.

#### 3) Implementation Detail

The optimization stage number $k$ is set to 9, which shows the best SSR effect among the following tests. The learning rate is set to 0.001, and the gradient-based optimization algorithm based on adaptive estimates of low-order moments (Adam [51]) is employed to train HSRnet. The trade-off parameter $\alpha$ for the loss function is set to 0.0001. The models are trained by Pytorch framework running in the Windows 10 environment with 16 GB RAM and one Nvidia RTX 2080 GPU.

#### 4) Experimental Dataset

The proposed HSRnet is evaluated by using the HSIs from CAVE [48] and remote sensing datasets.

##### a) CAVE dataset

CAVE dataset, which comprises 32 scenes with a size of 512 × 512, is a popular HSI dataset in HSI processing. All the HSIs in CAVE dataset cover the spectral range from 400 nm to 700 nm with a 10 spectral resolution containing 31 bands. Moreover, the RGB images covering the same scene as HSI data are available.

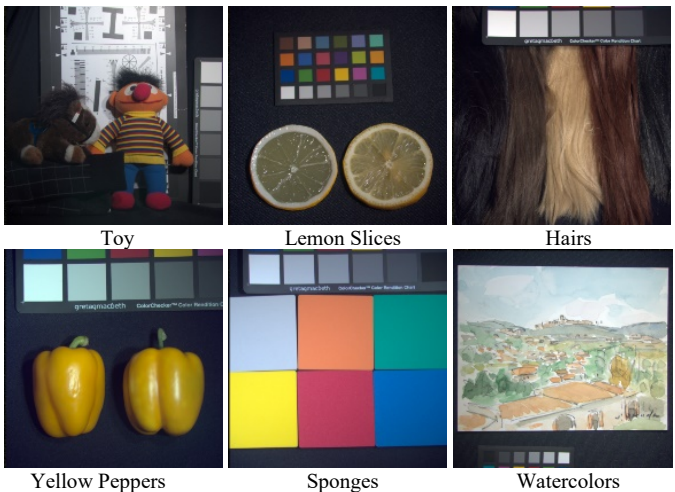

Toy Lemon Slices Hairs
Yellow Peppers Sponges Watercolors
Fig. 7 Six test images selected randomly in CAVE dataset

A total of 26 HSIs and the corresponding RGB images are randomly selected to prepare the training samples, and each image is split into 16 patches with a size of 128 × 128. Data augmentation is employed in this experiment because the insufficient training data are unfavorable to model training. The original training samples are flipped and rotated to increase the training data by eight times. The six remaining images are utilized for the test. The test images are shown in Fig. 7.

##### b) Remote Sensing dataset

**Sen2OHS dataset.** Images from four Chinese Orbita hyperspectral satellites (OHS) with 10 m spatial resolution are selected as HSIs to build a remote sensing dataset. OHS captures the HSIs in the spectral range from 400 nm to 1000 nm with 2.5 nm increments, but the HSI data sent to users are sampled to 32 bands.

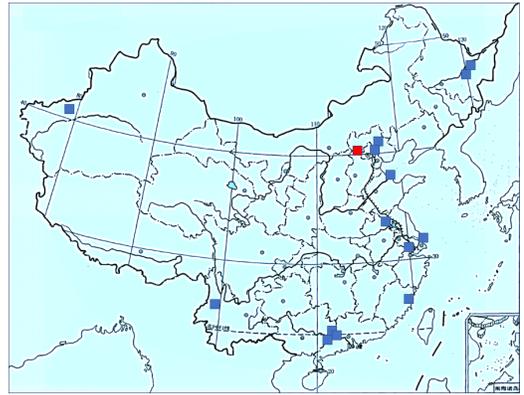

Fig. 8 Location of the training and testing images in Sen2OHS. The testing image is in red while training images are in blue.

The rich spectral information in OHS data with the 10 m spatial resolution is of considerable importance for application. However, free OHS data are mostly unavailable because of commerciality. This unavailability limits the hyperspectral data sources for researchers. Meanwhile, some MS images, such as Sentinel-2 bands with the same spatial resolution as OHS data (bands 2, 3, 4, and band 8), are available for free. Thus, Sen2OHS dataset is simulated to evaluate the SSR effect of the proposed model on the remote sensing data.

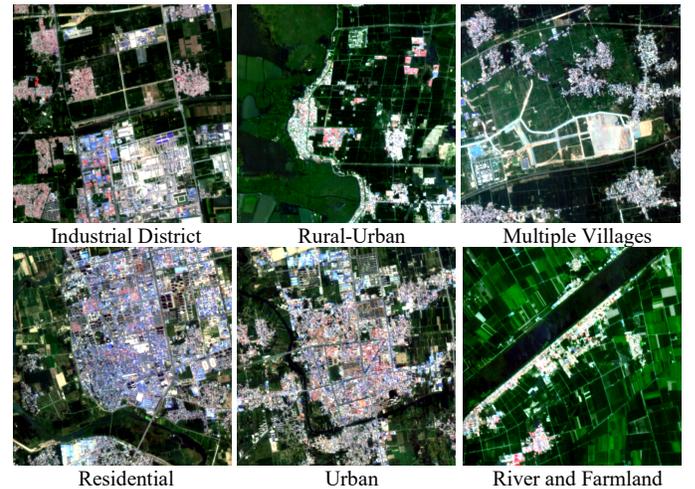

Industrial District Rural-Urban Multiple Villages
Residential Urban River and Farmland
Fig. 9 Six test images selected randomly in Sen2OHS dataset

Sentinel-2 MSIs are simulated from OHS HSIs by using Hysure [38] with the SRF of Sentinel-2 and OHS-A to reduce

the errors caused by geographic registration and the inconsistency of acquiring time between Sentinel-2 and OHS data. Furthermore, 6000 OHS HSIs with a size of 128 × 128 are selected for training from *the Competition in Hyperspectral Remote Sensing Image Intelligent Processing Application*[1]. The location of these images is shown in Fig. 8. And the testing images are randomly selected in Xiongan New Area, Hebei Province, China, as shown in Fig. 9.

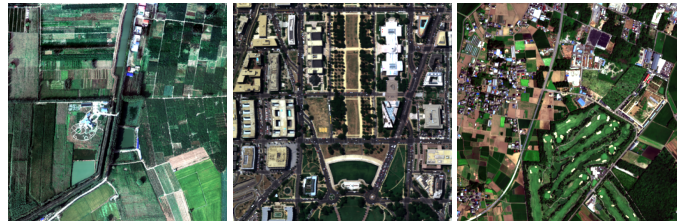

Xiongan          Washington DC Mall          Chikusei
Fig. 10  Three more HR remote sensing datasets

**HR Simulation dataset based on Sen2OHS.** Besides, to better verify the performance of models restoring spectral channels at different scales, three more datasets with a high spatial resolution are also simulated based on Sen2OHS, including Xiongan, Washington DC Mall, and Chikusei. Partial data of them are shown in Fig. 10. Xiongan dataset was an aerial image covered rural in Matiwan Village, Xiongan New Area, China, with a size of 3750×1580. The spectral range of Xiongan dataset is 400-1000 nm, with 250 bands and a spatial resolution of 0.5 meters. Washington DC Mall dataset [52] was acquired by HYDICE airborne sensor and with a size of 1280 ×307×210, covering the spectral wavelength from 400nm to 2500nm and the spatial resolution is lower than Xiongan and close to Chikusei. Chikusei dataset was taken by the Headwall Hyperspec-VNIR-C imaging sensor over agricultural and urban areas in Chikusei, Japan, with a size of 2517×2335 [53]. It contains 128 spectral bands ranging from 363 nm to 1018 nm with a spatial resolution of 2.5 meters. In the experiments, the spectral channels are downsampled to the same of OHS and Sentinel-2 by Hysure.

### B. Results on CAVE Dataset

#### 1) Quantitative and visual results

The quantitative results over six testing images are shown in Table IV, where the best results are in red bold and the second best is in blue. From the four quantitative image quality indexes, the deep learning-based methods show more remarkable amelioration in the spectral preservation than that in dictionary learning-based methods. Moreover, A+ performs well in spatial fidelity and is more highly improved compared with Arad. And the proposed HSRnet shows superior performance in spatial and spectral evaluation simultaneously.

In comparison to dictionary learning-based methods, the HSRnet achieves an average of 63.57% reduction in SAM and an average of 22.94% increase in PSNR. These findings illustrate that HSRnet can achieve effective spectral enhancement and maintain spatial information. Compared with other deep learning-based methods, HSRnet still shows some advantages in all indexes. HSCNN+ and sRCNN also show good spatial fidelity but get a worse spectral evaluation.

TABLE II
NUMERICAL COMPARISON OF FOUR QUANTITATIVE IMAGE QUALITY METRICS BETWEEN RESULTS ON CAVE DATASET

| Methods | CC | PSNR | SSIM | SAM |
|---|---|---|---|---|
| Arad | 0.9486 | 24.4613 | 0.7913 | 21.3129 |
| A+ | 0.9873 | 32.8830 | 0.9297 | 20.5403 |
| DenseUnet | 0.9907 | 32.5510 | 0.9642 | 8.1915 |
| CanNet | 0.9925 | 33.5975 | 0.9685 | 8.6435 |
| HSCNN+ | 0.9934 | 34.4354 | 0.9766 | 7.8048 |
| sRCNN | 0.9916 | 34.3669 | 0.9731 | 9.0175 |
| **HSRnet** | **0.9935** | **34.4903** | **0.9771** | **7.6208** |

Difference maps (DMs) between the reconstruction results and the ground truth are constructed to evaluate the results intuitively, as shown in Fig. 11. Six channels with wavelengths of 450, 500, 550, 600, 650, and 700 nm are selected. Fig. 11 demonstrates that Arad's result shows poor performance in spatial detail, as indicated in the background and the lemon pulp among all the presented bands. A+ obtains a better effect compared with that of Arad and even better than DenseUnet at some bands, such as 450 and 550 nm. CanNet shows a high difference in the edges. HSCNN+ can get good performance in several bands. But, HSRnet obtains DMs with the lowest value, which indicates that HSRnet achieves the best performance in SSR. As seen in DMs, HSRnet can adaptively accomplish spectral enhancement of different targets on the palette or the lemon slice. All methods perform poorly at the wavelength of 700 nm because of the insufficient spectral information.

#### 2) Discussion on fake and real lemon slices

Because there are fake and real lemon slices in the testing images, the reconstruction effects of methods at fake and real lemon slices are also presented. As shown in Fig. 12, the reflectance of real and fake lemon slices completely varies among bands 15to 31, namely the wavelength from 540 nm to 700 nm. The spectral curve of real lemon still increases after band 15. However, the spectral curve of fake lemon initially drops and then rises. In this case, deep learning-based methods can adaptively reconstruct the spectral detail of fake and real lemon slices separately. This reconstruction benefits from the powerful learning capability of CNNs, but Arad and A+ show poor performance in these bands. Although other deep learning-based algorithms can achieve good performance on distinguishing the spectrum of fake and real objects, the results of HSRnet show the highest similarity to the ground truth.

### C. Results on Remote Sensing Dataset

The proposed model is also verified on the remotely sensed dataset. Furthermore, four quantitative image quality indexes are employed to evaluate experimental results in the simulated experiments, including Sen2OHS dataset and three HR simulation datasets. After the simulated experiments, the trained model will be utilized to enhance the spectral resolution of real Sentinel-2 data. Moreover, a classification is presented to demonstrate the reliability of the reconstructed HSIs.

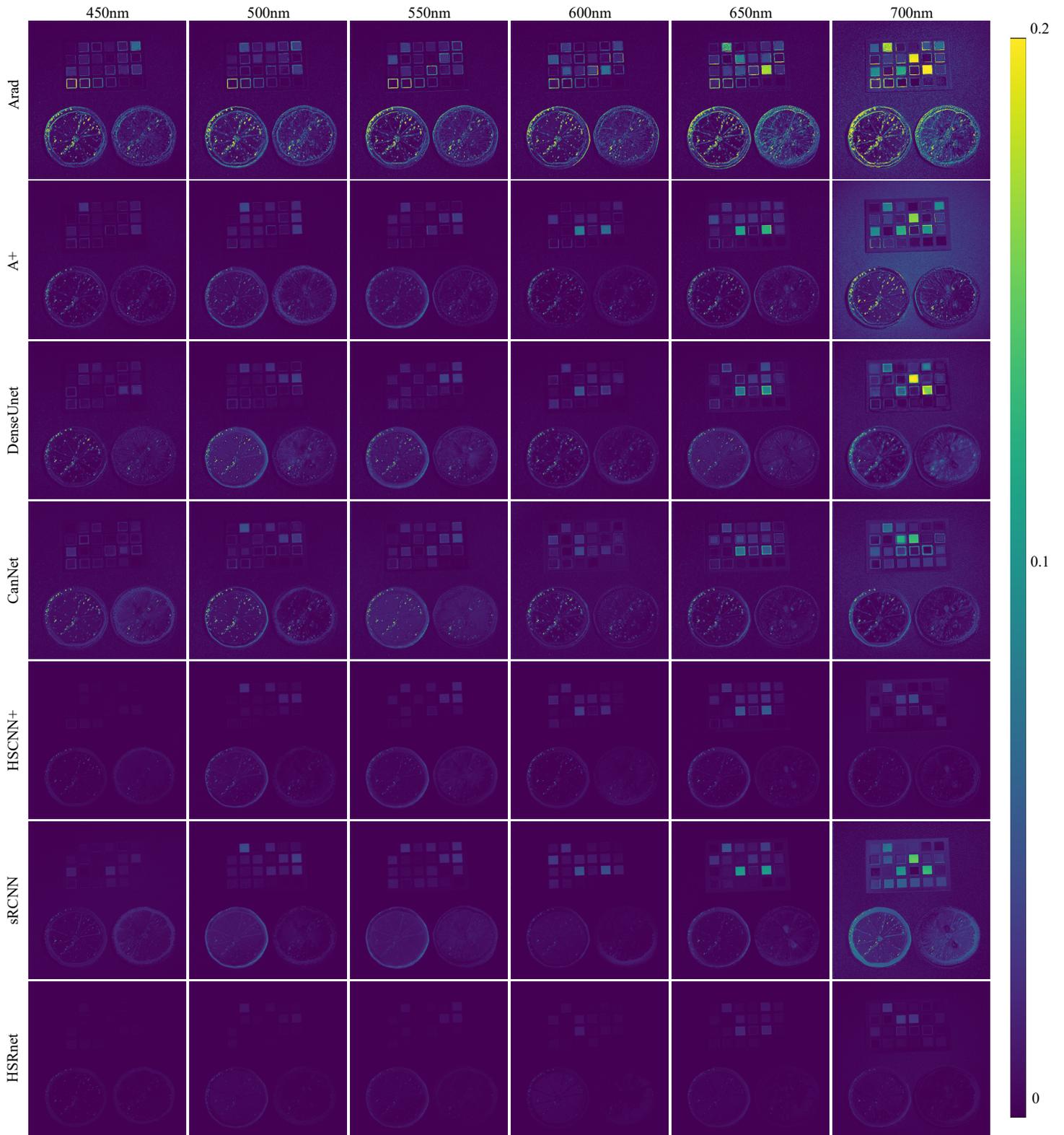

Fig. 11  Absolute differences of "Lemon Slices" image from CAVE dataset. Along 450, 500, 550, 600, 650, and 700 nm, the absolute differences between the reconstructed images and the ground truth are given. Each row from top to bottom is the result of Arad, A+, DenseUnet, CanNet, HSCNN+, sRCNN, and the proposed HSRnet.

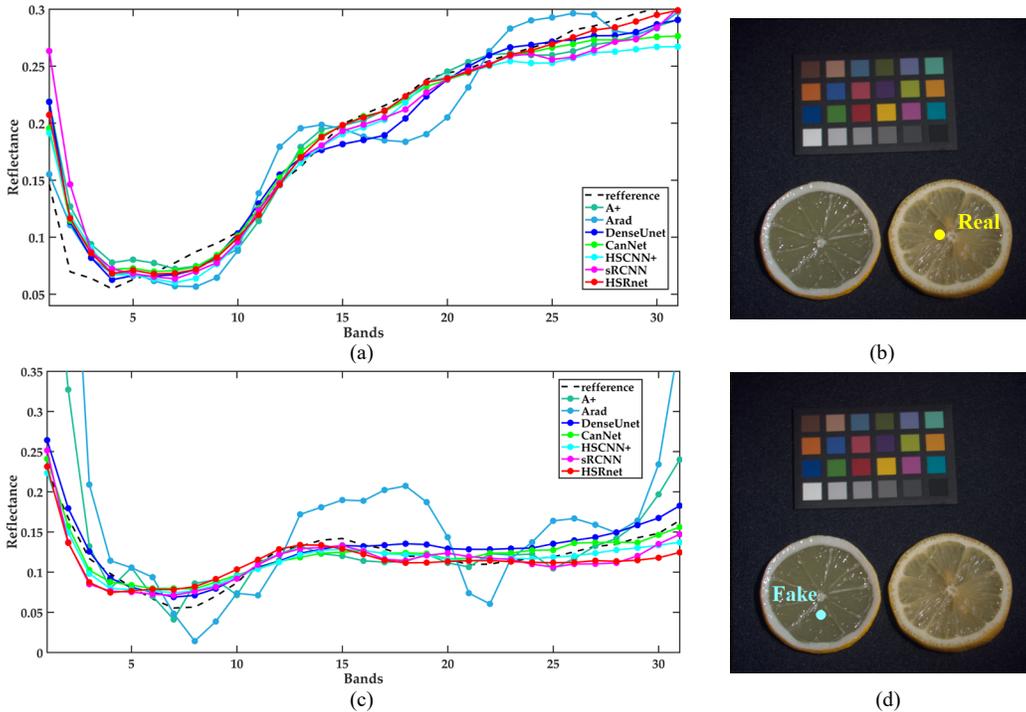

Fig. 12  Reflectance of "Lemon Slices" images from CAVE dataset at the fake and real lemon locations. (a) The reflectance at the real lemon slice location. (b) The real lemon slice location. (c) The reflectance at the fake lemon slice location. (d) The fake lemon slice location

### 1) Quantitative and visual results

#### a)  Sen2OHS Dataset

Table III shows the quantitative assessment results of testing images in Sen2OHS dataset. In contrast to the natural images, targets in remote sensing images are various and complex, resulting in poor spatial fidelity for all methods. The spectral preservation is improved because of the less color variation between targets than natural images. A+ and Arad show a sharp decline in CC, PSNR, and SSIM, which indicates a poor generalization effect. It's noted that the training samples of Arad and A+ are the same as those of deep learning-based methods, which are not divided into different domains unlike that of CAVE dataset, because effective models should be able to reconstruct images in different scenes adaptively with unified training samples.

TABLE III
NUMERICAL COMPARISON OF FOUR QUANTITATIVE IMAGE
QUALITY METRICS BETWEEN METHODS IN SEN2OHS
DATASET

| Methods | CC | PSNR | SSIM | SAM |
|---|---|---|---|---|
| Arad | 0.8149 | 22.4581 | 0.5631 | 11.0670 |
| A+ | 0.8592 | 24.4238 | 0.6924 | 9.5847 |
| DenseUnet | 0.9498 | 26.7262 | 0.8769 | 8.3135 |
| CanNet | 0.9621 | 28.1981 | 0.8901 | 7.4233 |
| HSCNN+ | 0.9593 | 28.8117 | 0.9164 | 6.9076 |
| sRCNN | 0.9689 | **29.2940** | **0.9389** | **6.5788** |
| HSRnet | **0.9725** | 28.9801 | 0.9344 | 6.8410 |

The proposed HSRnet improves the average CC, PSNR, and SSIM value by 16.18%, 23.63%, and 48.85%, respectively, compared with Arad and A+. The improvement of SAM is beyond 33.75%. Compared with four deep-learning methods, HSRnet shows a certain advantage in both spatial fidelity and spectral preservation. Surprisingly, sRCNN gain a tiny advantage over HSRnet, which is benefited by the spectra-by-spectra band optimization with huge computation.

The DMs of the selected testing image named "Urban" is shown in Fig. 13. Six bands, including bands 5, 10, 15, 20, 25, and 30, are displayed. The "Urban" image comprises rivers, farmlands, buildings, and other features, providing a considerable challenge to spectral SR. From the DMs of band 30, the spectral enhancement of farmlands with regular geometric shapes but diverse color brightness is difficult for dictionary learning-based methods. However, the sporadic buildings obtain improved spectral fidelity in A+ and Arad. For deep learning-based methods, with strong learning capability of different features, recovering the target with regular geometric shapes is easy, such as farmlands, streets, and rivers. But the recovery of various buildings, as shown in the results of deep learning-based methods on band 20, 30, shows unsatisfactory effect. This may be due to the inconsistently different geometric shapes of the same ground feature, which confuses CNN and mistakes them as different features. However, the results of the proposed HSRnet show lower error and less detail loss. Although sRCNN gets the best quantitative indexes, HSRnet shows more balanced visual results in different bands.

Fig. 14 (a) shows the average error of compared methods. The curve trend indicates that the spectral SR effect of the bands at the edge of the spectral coverage is worse than that of other bands. This finding has also been verified on CAVE dataset, which is due to the limited spectral information of bands at the edge of the spectral range obtained from the input multispectral images. Furthermore, all the compared methods yield slightly worse results on bands 9 to 21, as framed by magenta. As shown in Fig. 14 (b), the spectral range of Sentinel-2 and OHS-A is incompletely covered. The bands 9 to 12, 16 to 20, and 30 to 32 of the OHS data are not covered by Sentinel-2 SRF, thus yielding bands with poor spectral fidelity. However, with the SRF as a guide, HSRnet has good spectral reconstruction capability when spectral information of the relevant bands is deficient, which is shown in the figure with lower average errors as the form of fluctuations instead of a surge.

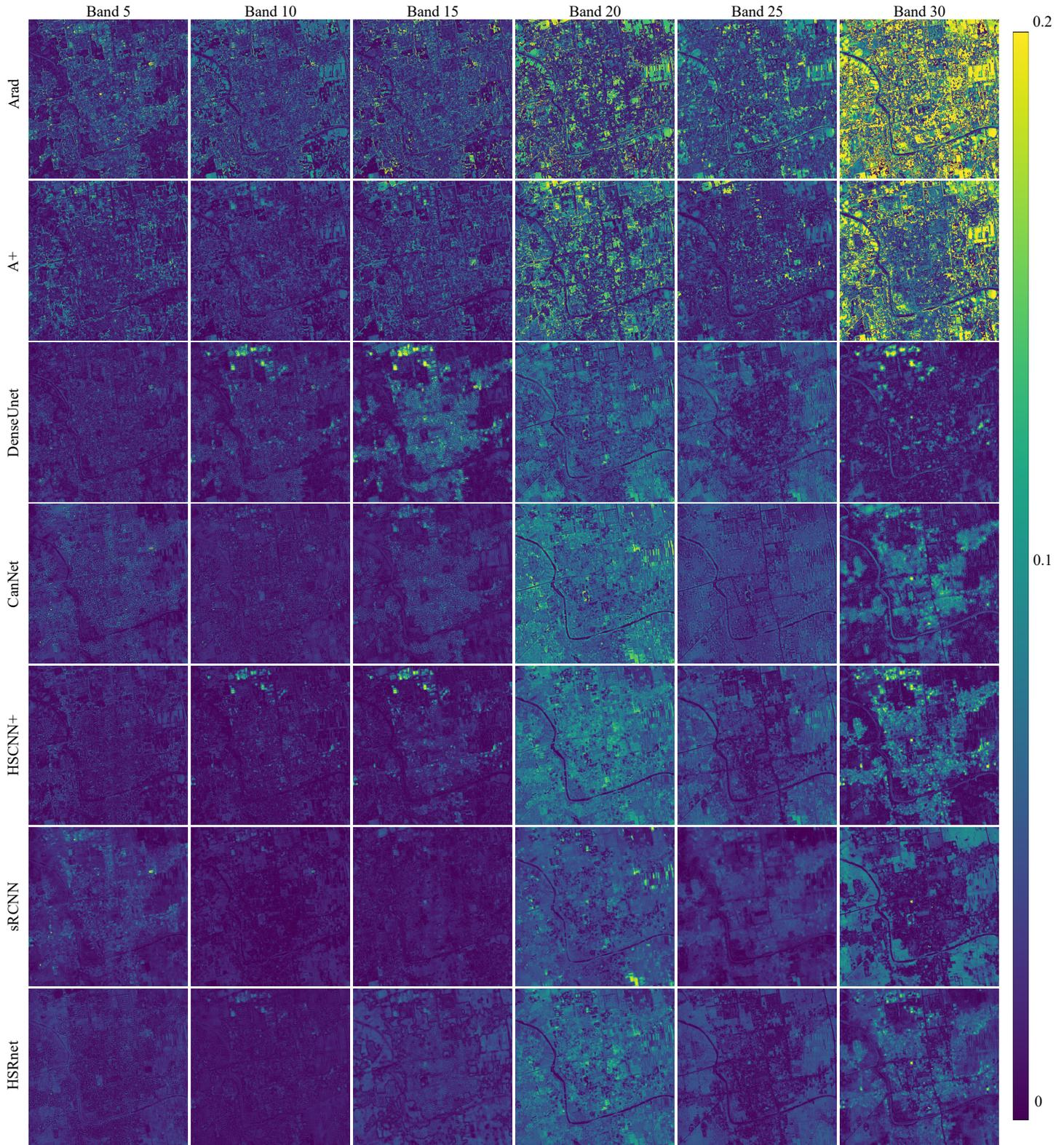

Fig. 13 Absolute differences of "Urban" image from Sen2OHS dataset. Along with bands 5, 10, 15, 20, 25, and 30, the absolute differences between the reconstructed images and the ground truth are given. Each row from top to bottom is the result of Arad, A+, DenseUnet, CanNet, HSCNN+, sRCNN, and the proposed HSRnet.

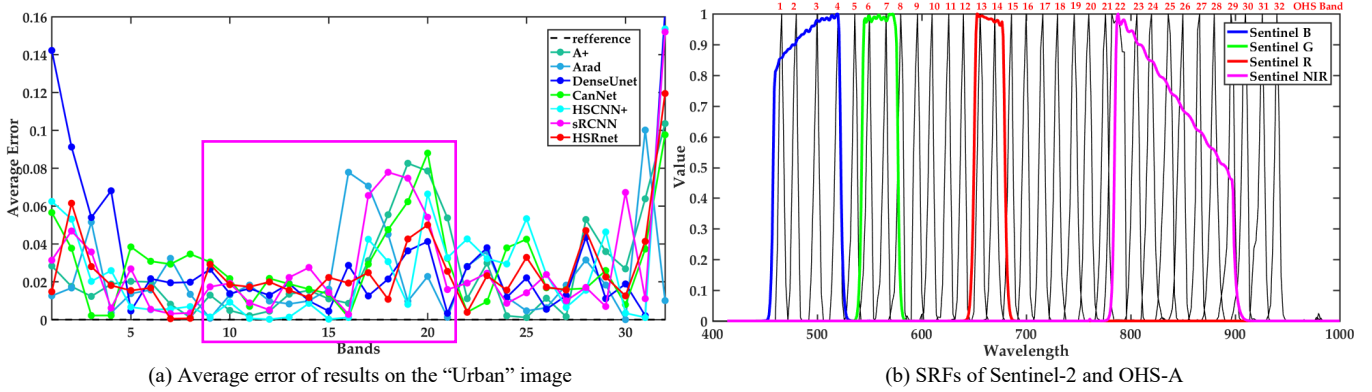

(a) Average error of results on the "Urban" image

(b) SRFs of Sentinel-2 and OHS-A

Fig. 14 (a) Average error of the results on the "Urban" image. Framed by magenta, results on band 9 to 21 of all methods show a poor tendency because of limited related spectral information in the input Sentinel-2 data. (b) The different SRFs of Sentinel-2 and OHS-A, which are used to help explain the phenomenon presented in (a).

### b) HR Simulation Datasets

To further compare the effect of the deep learning-based methods in different scales, synthetic datasets with finer spatial details, including Xiongan, Washington DC Mall, and Chikusei, are utilized. Quantitative results are shown in Table IV.

On these HR datasets, spectral superresolution becomes easier to achieve. With purer spectral information in HR training samples, deep learning-based methods can recover spectra more accurately, as shown in the table. And sRCNN shows good applicability in remote sensing datasets because of spectra-by-spectra optimization steps. Compared with sRCNN, the proposed HSRnet can get better performance with lower computational complexity. With the results of datasets at four different scales in remote sensing, the proposed HSRnet shows great stability and superiority than other deep learning-based algorithms in spectral fidelity.

| Dataset | Method | CC | PSNR | SSIM | SAM |
|---|---|---|---|---|---|
| Xiongan | DenseUnet | 0.9847 | 42.4634 | 0.9814 | 0.9217 |
| | CanNet | 0.9946 | 48.3492 | 0.9950 | 0.8029 |
| | HSCNN+ | 0.9942 | 48.4972 | 0.9959 | 0.7888 |
| | sRCNN | 0.9954 | 49.8814 | 0.9973 | 0.7623 |
| | HSRnet | **0.9963** | **50.7362** | **0.9973** | **0.7196** |
| Washington DC Mall | DenseUnet | 0.9927 | 39.7343 | 0.9848 | 1.8808 |
| | CanNet | 0.9987 | 47.8736 | 0.9971 | 1.1805 |
| | HSCNN+ | 0.9986 | 47.5770 | 0.9972 | 1.0983 |
| | sRCNN | 0.9989 | 48.5363 | 0.9978 | 1.0179 |
| | HSRnet | **0.9992** | **50.4457** | **0.9983** | **0.9395** |
| Chikusei | DenseUnet | 0.9897 | 39.2096 | 0.9809 | 4.0650 |
| | CanNet | 0.9967 | 44.2579 | 0.9933 | 3.6732 |
| | HSCNN+ | 0.9947 | 42.5542 | 0.9908 | **3.4254** |
| | sRCNN | 0.9955 | 43.4017 | 0.9924 | 3.5490 |
| | HSRnet | **0.9968** | **44.7133** | **0.9941** | 3.4528 |

#### 2) Classification results on real data

Owing to the good performance demonstrated on the remote sensing dataset, the trained HSRnet model is used on real Sentinel-2 data with 10 m spatial resolution to verify the reliability of the increased spectral information compared with the original MSI. We choose the classification experiments to evaluate it. The image is selected in the south of Nantes, France with a size of $512 \times 512$. The comparison results are shown in Fig. 15. The HSI is displayed with bands 27, 13, and 8 and the real Sentinel-2 data is shown with band 8, 4, and 3, where the vegetation is red.

The features are classified into 16 classes by using the support vector machine (SVM) as shown in the legend in Fig. 15. Additional spectral information is introduced to help combine the adjacent similar objects, and the classification results of the reconstructed HSI show less discrete objects.

| | OA | Kappa |
|---|---|---|
| Original MSI | 70.74% | 0.6296 |
| Reconstructed HSI | 73.22% | 0.6619 |

The quantitative evaluation also shows the increased spectral information recovered by HSRnet can help classification as presented in Table V. The classification results demonstrate improvements in OA and Kappa due to the additional spectral information in the reconstructed HSI. This improvement indicates that the proposed SSR method can accurately recover spectral information on the real dataset.

### D. Discussion

This section discusses the reliability of the proposed HSRnet, including ablation study and computational speed analysis.

#### 1) Ablation Study

The efficiency of the strategies of the proposed HSRnet, including optimization stages, parametric self-learning based on channel attention module, SRF-guided initial restoration network, and fast joint spatial-spectral loss, is first discussed as shown in Table VI. A 19-layer Resnet [24] is chosen as a baseline. OS, CAM, SRF, and SAM Loss represent the aforementioned strategies, and the details will be provided later.

**Optimization Stages.** Compared with Resnet, HSRnet with only optimization stages (namely, HSRnet w/o CAM in Table VI) shows substantially high superiority in spatial and spectral fidelity. Compared with DenseUnet, the proposed network with physical interpretability shows a slight advantage without the help of other strategies.

**Channel Attention Module.** Comparing HSRnet without SRF with HSRnet without CAM, HSRnet with parametric self-learning based on channel attention module shows improved spatial fidelity and spectral enhancement due to the capability to learn parameters adaptively for different iterations and bands.

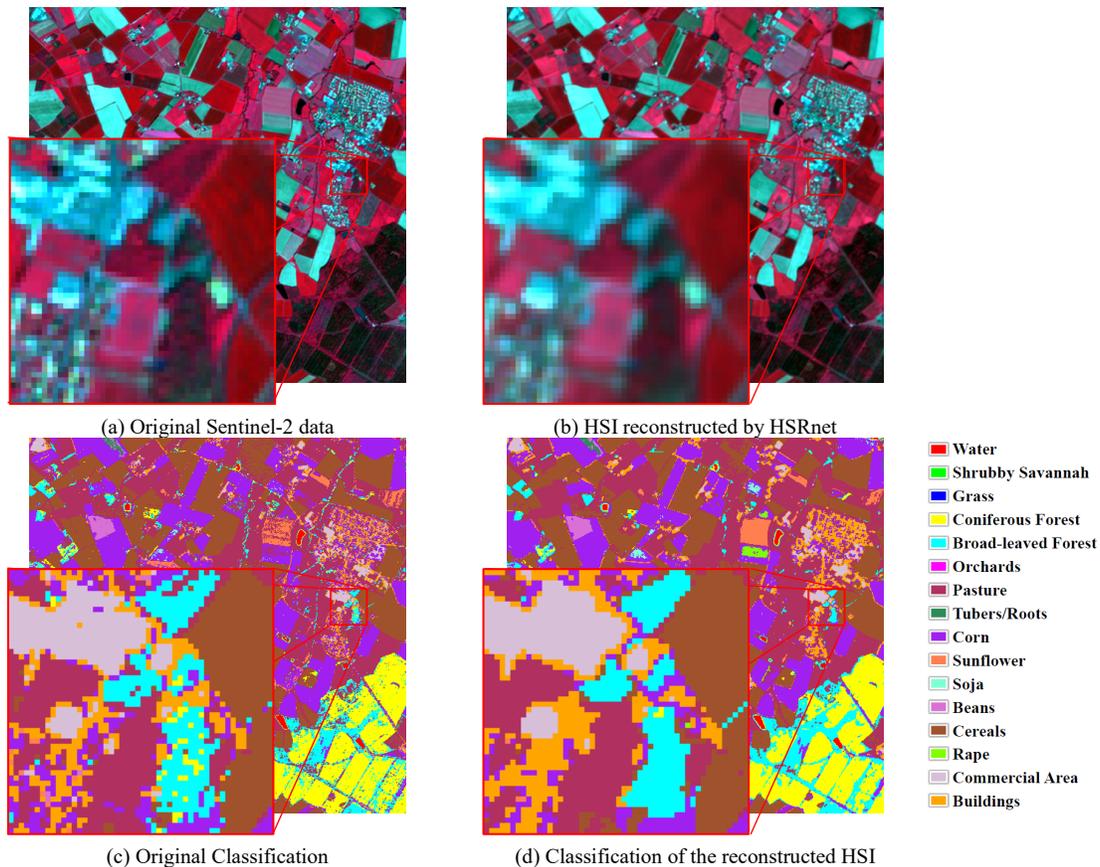

(a) Original Sentinel-2 data

(b) HSI reconstructed by HSRnet

| | Water |
| :---: | :--- |
| | Shrubby Savannah |
| | Grass |
| | Coniferous Forest |
| | Broad-leaved Forest |
| | Orchards |
| | Pasture |
| | Tubers/Roots |
| | Corn |
| | Sunflower |
| | Soja |
| | Beans |
| | Cereals |
| | Rape |
| | Commercial Area |
| | Buildings |

(c) Original Classification

(d) Classification of the reconstructed HSI

Fig. 15  Classification comparison on the real Sentinel-2 data and the reconstructed HSI by HSRnet.

TABLE VI
ABLATION STUDY OF THE PROPOSED STRATEGIES ON CAVE DATASET

| | OS | CAM | SRF | SAMLoss | CC | PSNR | SSIM | SAM |
| :--- | :---: | :---: | :---: | :---: | :---: | :---: | :---: | :---: |
| ResNet | × | × | × | × | 0.9843 | 28.4483 | 0.9415 | 11.4720 |
| DenseUnet | - | - | - | - | 0.9907 | 32.5510 | 0.9642 | 8.1915 |
| HSRnet w/o CAM | √ | × | × | × | 0.9919 | 33.3288 | 0.9674 | 8.2279 |
| HSRnet w/o SRF | √ | √ | × | × | 0.9930 | 34.2748 | 0.9741 | 8.0927 |
| HSRnet w/o SAMLoss | √ | √ | √ | × | 0.9933 | 34.3467 | 0.9742 | 7.8506 |
| HSRnet | √ | √ | √ | √ | **0.9935** | **34.4903** | **0.9771** | **7.6208** |

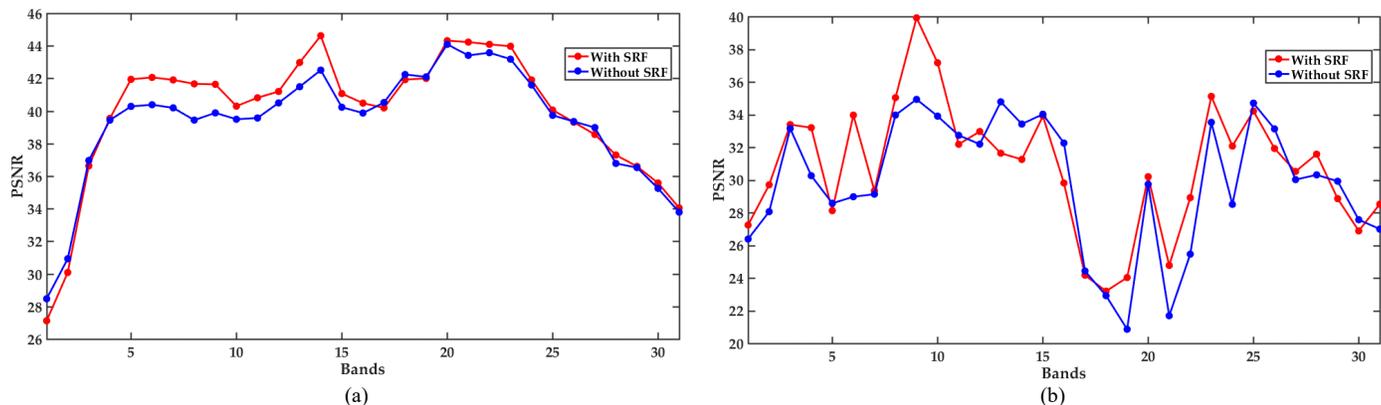

(a)

(b)

Fig. 16  PSNR of results reconstructed with SRF as a guide and without SRF on two datasets. (a) Comparison on CAVE dataset. (b) Comparison on Sen2OHS dataset.

**SRF-guided IRN.** With the SRF as a guide, HSRnet shows tiny spatial improvements but substantially good spectral maintaining as shown in the values of SAM (i.e. HSRnet w/o SAMLoss in Table VI). Furthermore, the comparison with HSRnet without SRF on CAVE dataset is shown in Fig. 16. As shown in Fig. 16 (a), the PSNR of results reconstructed by the model with SRF as a guide is higher than the model without an SRF guide. This finding shows that the SRF guide can help HSRnet achieve improved performance. Fig. 16 (b) shows the same conclusion on the remote sensing dataset.

**Spatial-Spectral Loss Function.** With SAMLoss, the proposed HSRnet shows some improvement not only on SAM but also on other metrics of spatial fidelity. It states that considering the spectral loss with spatial loss function, the spatial fidelity and spectral preservation can be mutually reinforced.

### 2) Computational Speed Analysis

Deep learning-based methods can achieve satisfying spectral enhancement on CAVE and Sen2OHS datasets due to their strong non-linear mapping capability, and the parameter number is very important to them. For example, as the parameter number increases, CNN can reach effective performance without changing the structure by computing additional features in convolution layers. Thus, the comparison between deep learning-based methods in parameter number and running time is performed with similar feature numbers.

optimization stages in HSRnet. Besides, CanNet owns the least parameters because it works as a shallow network. Although the parameter number in DenseUnet are approximately twice as many as the proposed HSRnet in total, HSRnet shows better performance in SSR compared with DenseUnet. FLOPs show the algorithm complexity by floating-point operations. With pixel-by-pixel optimization, sRCNN gets the highest FLOPs although the parameter number is similar to HSRnet, which leads to a long running time. DenseUnet benefits from the down- and upsampling to get the fewest FLOPs. Although DenseUnet can train an epoch faster than HSRnet, it converges at 200 epochs. Without downsampling to fast calculation, HSRnet spends more training time in each epoch but converges earlier than that of other networks, as shown in Fig. 17.

TABLE VII
COMPUTATIONAL SPEED ANALYSIS OF DEEP LEARNING-BASED METHODS ON CAVE DATASET

|  | DenseUnet | sRCNN | CanNet | HSCNN+ | HSRnet |
|---|---|---|---|---|---|
| Params | 1360.1K | 789.3K | 163.0K | 915.1K | 769.7K |
| FLOPs | $3.02\times10^{10}$ | $5.96\times10^{12}$ | $3.97\times10^{10}$ | $2.23\times10^{11}$ | $1.79\times10^{11}$ |
| Training | 68655s | 146539s | 49285s | 57805s | 30831s |
| Test | 1.2598s | 4.5950s | 1.2387s | 1.7996s | 1.5364s |

Tables VII lists the parameter numbers, floating-point operations (FLOPs), training and test time of deep learning methods. Training and test time are all counted on CAVE dataset. DenseUnet obtains numerous parameters in down-and-up stages due to dense blocks, while most effort is put into

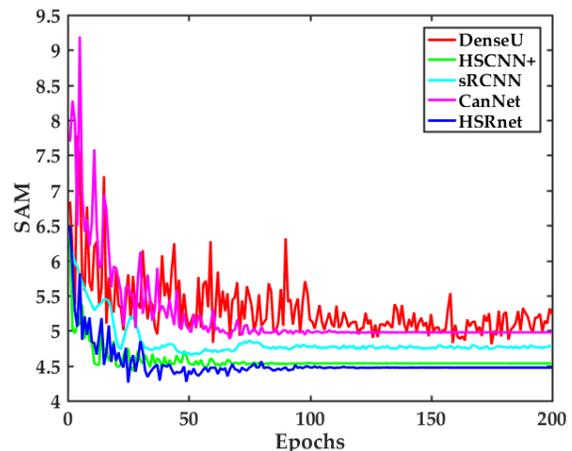

Fig. 17 Validation loss of deep learning-based methods

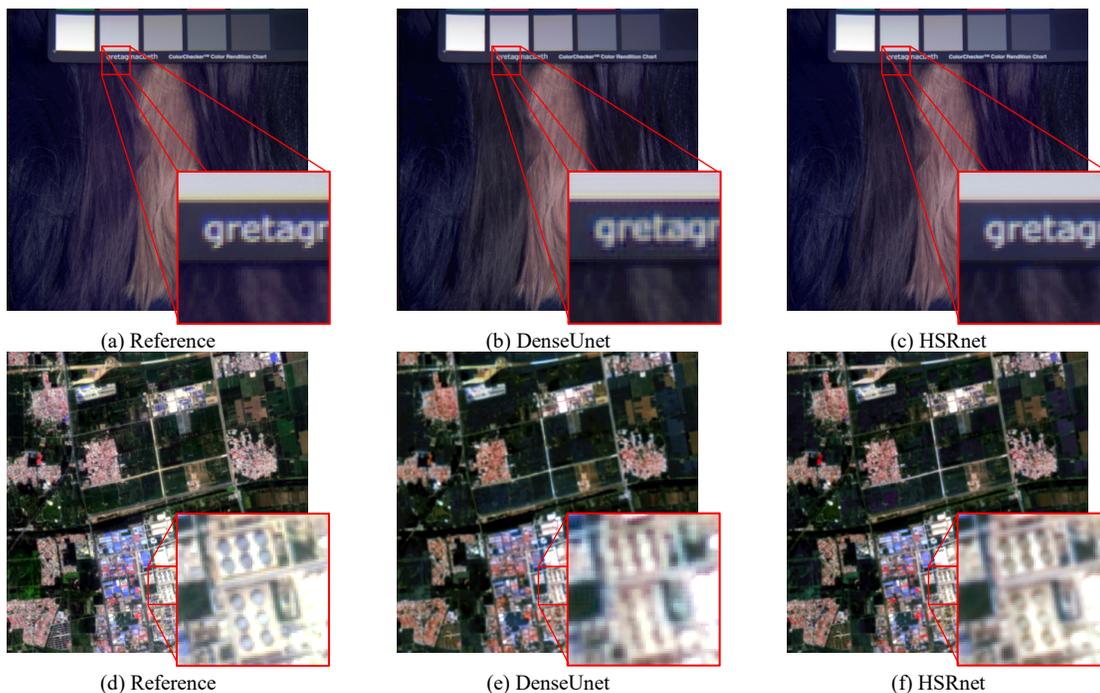

(a) Reference     (b) DenseUnet     (c) HSRnet

(d) Reference     (e) DenseUnet     (f) HSRnet

Fig. 18 Partially enlarged details of the results on "Hair" and "Industrial District" image. (a) The reference "Hair" image in CAVE dataset shown by bands 14, 7, and 2. (b) Result of DenseUnet shown by the same band combination. (c) Result of HSRnet. (d) The ground truth of the "Industrial District" image in Sen2OHS dataset shown in bands 14, 7, and 2. (e) Result of DenseUnet shown by the same band combination. (f) Result of HSRnet.

As discussed above, DenseUnet can accelerate the calculation by downsampling the input images. However, this acceleration compromises spatial details, as shown in Fig. 18. Whether on CAVE or Sen2OHS dataset, DenseUnet shows spatial blurry effects, whereas HSRnet can maintain good spatial fidelity with rich details, such as the cylindrical buildings in Sen2OHS results and the clear letter edges in CAVE results. Furthermore, the HSRnet results suffer from mild spatial degradation on Sen2OHS dataset. Notably, the spatial resolution of the captured OHS-A data is not accurately 10 m. This value is slightly coarser than that of Sentinel-2, resulting in spatial degradation.

The proposed HSRnet owns acceptable parameter numbers and computation complexity but gets the best SSR performance. Furthermore, considering the effect and running time, HSRnet maintains more spatial details with fewer parameters and acceptable test time. In addition, HSRnet realizes early convergence, although the training time of HSRnet is longer than that of other methods in one iteration, resulting in less total training time. Thus, a conclusion can be drawn that building CNN with physical logic is superior to using data-driven CNN.

## IV. CONCLUSION

This paper presents an SRF-guided optimization-driven SSR network with spatial-spectral prior to enhance the spectral information of the MS/RGB image. The traditional gradient descent-based algorithm is transformed into an end-to-end CNN with the help of deep spatial-spectral prior. The proposed HSRnet groups the spectral similar bands using the physical information and the spectral response function to reconstruct different spectral ranges instead of the traditional black-box data-driven CNN. Using the CAM blocks to learn parameters rather than a manual setting can automatically adjust the weights for different channels rather than a fixed value to the entire image. Moreover, the proposed HSRnet transforms the optimization model into a data-driven model. This model provides CNN with physical interpretability and facilitates flexible learning of optimization parameters in an end-to-end manner. Experimental results on natural and remotely sensed datasets confirm the feasibility and superiority of the proposed method. Furthermore, as shown in both datasets, especially in Sen2OHS dataset, the spectral coverage between input and output data plays an important role in the model effect. Thus, the effective utilization of MS bands with different spatial resolutions to reach complete coverage of spectral information and achieve spatial-spectral SR is a direction of our future works.